\documentstyle[aps,pra,floats,epsfig]{revtex}

\newcommand{\be}{\begin{equation}}
\newcommand{\ee}{\end{equation}}
\newcommand{\bea}{\begin{eqnarray}}
\newcommand{\eea}{\end{eqnarray}}
\newcommand{\bt}{\begin{tabular}}
\newcommand{\et}{\end{tabular}}
\newcommand{\ba}{\begin{array}}
\newcommand{\ea}{\end{array}}

\begin{document}

\twocolumn[\hsize\textwidth\columnwidth\hsize\csname
@twocolumnfalse\endcsname

\small{\hfill{
\begin{tabular}{l}
DSF$-$6/2001 \\
SISSA$-$12/2001/EP \\
physics/0102020
\end{tabular}}}


\title{On a universal photonic tunnelling time}
\author{Salvatore Esposito$^\ast$}

\address{Dipartimento di Scienze Fisiche, Universit\`{a} di Napoli
``Federico II''\\ and \\
Istituto Nazionale di Fisica Nucleare, Sezione di Napoli \\
Complesso Universitario di Monte S. Angelo,  Via Cinthia, I-80126 Napoli,
Italy \\
$^\ast$Visiting scientist at SISSA--ISAS, Via Beirut 4, I-34013 Trieste,
Italy \\
E-mail: Salvatore.Esposito@na.infn.it }

\maketitle

\begin{abstract}
We consider photonic tunnelling through evanescent regions and obtain
general analytic expressions for the transit (phase) time $\tau$ (in the
opaque barrier limit) in order to study the recently proposed
``universality'' property according to which $\tau$ is given by the
reciprocal of the photon frequency. We consider different physical
phenomena (corresponding to performed experiments) and show that such a
property is only an approximation. In particular we find that the
``correction'' factor is a constant term for total internal reflection and
quarter-wave photonic bandgap, while it is frequency-dependent in the case
of undersized waveguide and distributed Bragg reflector. The comparison of
our predictions with the experimental results shows quite a good agreement
with observations and reveals the range of applicability of the
approximated ``universality'' property.
\end{abstract}


\vskip2pc]


\section{Introduction}

\noindent
In recent times, some photonic experiments \cite{under}-\cite{haibel}
dealing with evanescent mode propagation have drawn some attention because
of their intriguing results. All such experiments have measured the time
required for the light to travel through a region in which only evanescent
propagation occurs, according to classical Maxwell electrodynamic. If
certain conditions are fulfilled (i.e. in the limit of opaque barriers),
the obtained transit times are usually {\it shorter} than the corresponding
ones for real (not evanescent) propagation through the same region. Due to
the experimental setups, this has been correctly interpreted in terms of
group velocities \cite{group} greater than $c$ inside the considered
region. Although there has been some confusion in the scientific community,
leading also to several different definitions of the transit time
\cite{times}, these results are not at odds with Einstein causality since,
according to Sommerfeld and Brillouin \cite{sommbrill}, the front velocity
rather than the group velocity is relevant for this. Waves which are
solutions of the Maxwell equations always travel in vacuum with a front
velocity equal to $c$ while, in certain conditions, their phase and group
velocities can be different from $c$ \cite{espo}. It is worthwhile to
observe that the quoted experiments are carried out studying different
phenomena (undersized waveguide, photonic bandgap, total internal
reflection) and exploring different frequency ranges (from optical to
microwave region). \\ The interest in such experiments is driven by the
fact that evanescent mode propagation through a given region can be viewed
as a photonic tunnelling effect through a ``potential'' barrier in that
region. This has been shown, for example, in Ref. \cite{analogy} using the
formal analogy between the (classical) Helmholtz wave equation and the
(quantum mechanical) Schr\"{o}dinger equation (see also Ref. \cite{or}). In this respect, the photonic
experiments are very useful to study the question of tunnelling times,
since experiments involving charged particle (e.g. electrons) are not yet
sensible enough to measure transit times due to some technical difficulties
\cite{electrons}. \\ From an experimental point of view, the transit time
$\tau$ for a wave-packet propagating through a given region is measured as
the interval between the arrival times of the signal envelope at the two
ends of that region whose distance is $D$. In general, if the wave-packet
has a group velocity $v_g$, this means that $\tau = D / v_g$. Since $v_g =
d \omega /dk$ ($k$ wave-vector, $\omega$ angular frequency), then we can
write \cite{merz}:
\be  \label{1}
\tau \; = \; \frac{d \phi}{d \omega} ~~~,
\ee
where $d \phi \, = \, D \, dk$ is the phase difference acquired by the
packet in the considered region. The above argument works as well for
matter particles in quantum mechanics, changing the role of angular
frequency and wave-vector into the corresponding ones of energy and
momentum through the Planck - de Broglie relations. \\
However, difficulties arise when we deal with tunnelling times, since
inside a barrier region the wave-vector (or the momentum) is imaginary,
and hence no group velocity can be defined. As a matter of fact, different
definitions of tunnelling time exist. While we refer the read to the
quoted literature \cite{times}, here we use the simple definition of phase
time which coincides with Eq. (\ref{1}). In fact, although $v_g$ seems
meaningless in this case, nevertheless Eq. (\ref{1}) is meaningful also
for evanescent propagation. The adopted point of view takes advantage of
the fact that experimental results \cite{under}-\cite{haibel} seem to confirm
the definition of phase time for the tunnelling transit time. \\
Recently, Haibel and Nimtz \cite{haibel} have noted that, regardless of
the different phenomena studied, all experiments have measured photonic
tunnelling times which are approximately equal to the reciprocal of the
frequency of the radiation used in the given experiment. Such a
``universal'' behaviour is quite remarkable in view of the fact that,
although photonic barrier traversal takes place in all the quoted
experiments, nevertheless the boundary conditions are peculiar of each
experiment. \\
In the present paper we carefully study the proposed universality starting
from a common feature of tunnelling phenomena and, in the following
section, derive a general expression for the transit (phase) time.
Different experiments manifest themselves into different dispersion
relations for the barrier region. We then analyze each peculiar experiment
in Sects. III,IV,V and compare theoretical predictions with experimental
observations. Finally, in Sect. VI, we discuss our results and give
conclusions. \\
Note that, differently from other possible analysis (see, for example, the
comparison with a photonic bandgap experiment in \cite{scalora}), we deal
with only tunnelling times, which have been directly observed, and not
with velocities which, in the present case, are derived from transit
times.

\section{Phase time and dispersion relation}

\begin{figure}[t]
\epsfysize=4cm
\epsfxsize=6cm
\centerline{\epsffile{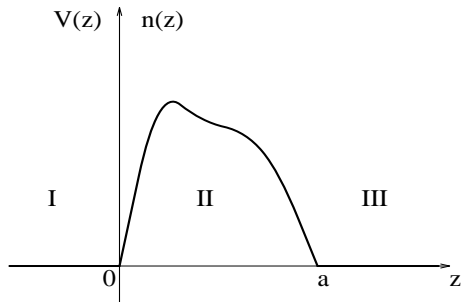}}
\caption{A barrier potential $V(z)$ for a particle or a barrier refractive
index $n(z)$ for an electromagnetic wave.}
\label{barrier}
\end{figure}

\noindent
In this paper we study one-dimensional problems or, more in general,
phenomena in which evanescent propagation takes place along one direction,
say $z$. Let us then consider a particle or a wave-packet moving along the
$z$-axis entering in a region $[0,a]$ with a potential barrier $V(z)$ or a
refractive index $n(z)$, as depicted in Figure \ref{barrier}. The
energy/frequency of the incident particle/wave is below the maximum of the
potential or cutoff frequency. For all experiments we'll consider, the
barrier can be modelled as a square one, in which $V(z)$ or $n(z)$ is
constant in regions I,II,III but different from one region to another. We
also assume that $V(z)$ or $n(z)$  is equal in I and III and take this
value as the reference one. \\
The propagation of the particle/wave through the barrier is described a by
a scalar field $\psi$ representing the Schr\"{o}dinger wave function in the
particle case or some scalar component of the electric or magnetic field
in the wave case. (The precise meaning of $\psi$ in the case of wave
propagation depends on the particular phenomenon we consider. However, the
aim of this paper is to show that a common background for all tunnelling
phenomena exist). Given the formal analogy between the Schr\"{o}dinger
equation and the Helmholtz equation \cite{analogy}, \cite{or}, this function takes
the following values in regions I,II,III, respectively:
\bea
\psi_{I} &=& e^{i k z} \, + \, R \, e^{- i k z}  \label{2} \\
\psi_{II} &=& A \, e^{- \chi z} \, + \, B \, e^{\chi z}  \label{3} \\
\psi_{III} &=& T \, e^{i k (z-a)}  \label{4} ~~~,
\eea
where $k$ and $k_2=i \chi$ are the wave-vectors ($p = \hbar k$ is the
momentum) in regions I (or III) and II, respectively. Note that we
have suppressed the time dependent factor $e^{i \omega t}$. Obviously, the
physical field is represented by a wave-packet with a given spectrum in
$\omega$:
\be
\psi(z,t) \; = \; \int \, d \omega \, \eta(\omega) \, e^{i (kz - \omega t)}
~~~.
\ee
where $\eta(\omega)$ is the envelope function. Keeping this in mind we use,
however, for the sake of simplicity, the simple expressions in Eqs.
(\ref{2}), (\ref{3}), (\ref{4}). Furthermore, for the moment, we disregard
the explicit expression fro $k$ and $\chi$ in terms of the angular
frequency $\omega$ (or the relation between $p$ and $E=\hbar \omega$). As
well known, the coefficients $R,T,A,B$ can be calculated from the matching
conditions at interfaces:
\bea
\psi_{I}(0) \; = \; \psi_{II}(0)  ~~~~&,&~~~~
\psi_{II}(a) \; = \; \psi_{III}(a) \label{5} \\
\psi_{I}^\prime(0) \; = \; \psi_{II}^\prime(0)  ~~~~&,&~~~~
\psi_{II}^\prime(a) \; = \; \psi_{III}^\prime(a) \label{6}  ~~~,
\eea
where the prime denotes differentiation with respect to $z$. Substituting
Eqs. (\ref{2}), (\ref{3}), (\ref{4}) into (\ref{5}), (\ref{6}) we are then
able to find $R,T,A,B$ and thus the explicit expression for the function
$\psi$. Here we focus only on the transmission coefficient $T$; its
expression is as follows:
\be  \label{7}
T \; = \; \left[ 1 \, - \, r^2 \, e^{-2 \chi a} \right]^{-1} \, \left( 1
\, - \, r^2 \right) \, e^{- \chi a}
\ee
with:
\be  \label{8}
r \; = \; \frac{\chi + i k}{\chi - i k} ~~~.
\ee
The interesting limit is that of opaque barriers, in which $\chi a \gg 1$.
All photonic tunnelling experiments have mainly dealt with this case, in
which ``superluminal'' propagation is predicted \cite{hartman}.
Taking this limit into Eq. (\ref{7}) we have:
\be  \label{9}
T \; \simeq  \; 2 \, \left[ 1 \, - \, i \, \frac{k^2 - \chi^2}{2 k \chi}
\right]^{-1} \, e^{- \chi a} ~~~.
\ee
The quantity $\phi$ in Eq. (\ref{1}), relevant for the tunnelling time, is
just the phase of $T$:
\be  \label{10}
\phi \; \simeq \; \arctan \, \frac{k^2 - \chi^2}{2 k \chi} ~~~.
\ee
The explicit evaluation of $\tau$ in Eq. (\ref{1}) depends, clearly, from
the dispersion relations $k = k(\omega)$ and $\chi = \chi (\omega)$.
However, by substituting Eq. (\ref{10}) into (\ref{1}) we are able to
write:
\be  \label{11}
\tau \; = \; 2 \, \left[ 1 \, + \, \left( \frac{k}{\chi} \right)^2
\right]^{-1} \, \frac{d~}{d \omega} \, \frac{k}{\chi} ~~~,
\ee
showing that $\tau$ depends only on the ratio $k/\chi$. We can also obtain
a particularly expressive relation by introducing the quantities:
\be  \label{12}
\frac{k_1}{v_1} \; = \; k \, \frac{dk}{d\omega} ~~~~~,~~~~~
\frac{k_2}{v_2} \; = \; - \, \chi \, \frac{d\chi}{d\omega} ~~~.
\ee
In fact, in this case we get:
\be   \label{13}
\tau \; = \; \frac{2}{\chi k} \, \left[ \frac{\chi^2}{k^2 + \chi^2} \,
\frac{k_1}{v_1} \, + \, \frac{k^2}{k^2 + \chi^2} \, \frac{k_2}{v_2}
\right] ~~~.
\ee
Note that while $k_1$ and $k_2$ are the real or imaginary wave-vectors in
regions I (or III) and II, $v_1$ and $v_2$ represent the ``real''
or ``imaginary'' group velocities in the same regions. Obviously, an
imaginary group velocity (which is the case for $v_2$) has no physical
meaning, but we stress that in the physical expression for the time $\tau$
in (\ref{13}) only the ratio $k_2/v_2$ enters, which is a well-defined
real quantity. \\
Equations (\ref{11}) and (\ref{13}) are very general ones (holding in the
limit of opaque barriers): they apply to {\it all} tunnelling phenomena.
It is nevertheless clear that peculiarities of a given experiment enter in
$\tau$ only through the dispersion relations $k = k(\omega)$ and $\chi =
\chi(\omega)$ or, better, $k(\omega)/\chi(\omega)$. \\
As an example of application of the obtained general formula, we here
consider the case of tunnelling of non relativistic electrons with mass
$m$ through a potential square barrier of height $V_0$. (In the next
sections we then study in detail the three types of experiment already
performed). The electron energy is $E= \hbar \omega$ (with $E<V_0$) while
the momenta involved in the problem are $p=\hbar k$ and $i q = \hbar k_2 =
i \hbar \chi$. In this case, the dispersion relations read as follows:
\bea
k &=& \sqrt{\frac{2 m \, \omega}{\hbar}}  \label{14} \\
\chi &=& \sqrt{\frac{2 m (V_0 - \hbar \omega)}{\hbar^2}}  \label{15}
\eea
and thus:
\be  \label{16}
\frac{k}{\chi} \; = \; \sqrt{\frac{\hbar \omega}{V_0 -
\hbar \omega}} ~~~.
\ee
By substituting into Eq. (\ref{11}) we immediately find:
\be   \label{17}
\tau \; = \; \frac{\hbar}{\sqrt{E (V_0 - E)}} \; = \; \frac{1}{\hbar} \;
\frac{2m}{\chi k} ~~~.
\ee
\begin{figure}[t]
\epsfysize=4.5cm
\epsfxsize=6cm
\centerline{\epsffile{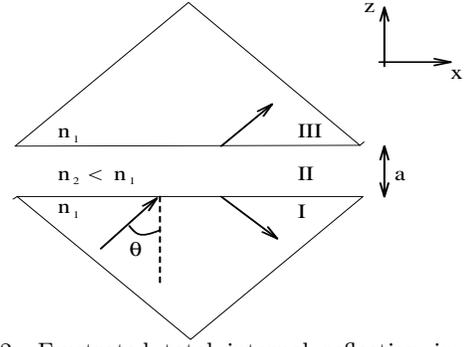}}
\caption{Frustrated total internal reflection in a double prism.}
\label{prism}
\end{figure}

\section{Total internal reflection}

\noindent
The first photonic tunnelling phenomenon we consider is that of frustrated
total internal reflection \cite{ghatak}. This is a two-dimensional
process, but tunnelling proceeds only in one direction. With reference to
Figure \ref{prism}, a light beam impinges from a dielectric medium
(typically a prism) with index $n_1$ onto a slab with index $n_2 < n_1$.
If the incident angle is greater than the critical value $\theta_c =
\arcsin n_2/n_1$, most of the beam is reflected while part of it tunnels
through the slab and emerges in the second dielectric medium with index
$n_1$. Note that wave-packets propagate along the $x$ direction, while
tunnelling occurs in the $z$ direction. \\
The wave-vectors  $k_1, k_2$ in regions I (or III) and II satisfy:
\bea
k_1^2 &=& k_x^2 \, + \, k^2  \label{18} \\
k_2^2 &=& k_x^2 \, - \, \chi^2  \label{19} ~~~,
\eea
where $k_x$ is the $x$ component of $k_1$ or $k_2$ and $k,\chi$ are as
defined in the previous section. The dispersion relations in regions I
(or III) and II are, respectively:
\bea
k_1 &=& \frac{\omega}{c} \, n_1  \label{20}  \\
k_2 &=& \frac{\omega}{c} \, n_2  \label{21}  ~~~.
\eea
These equations also define the introduced quantities:
\bea
v_1 &=& \frac{c}{n_1}  \label{22}  \\
v_2 &=& \frac{c}{n_2}  \label{23}  ~~~.
\eea
It is now very simple to obtain the tunnelling time in the opaque barrier
limit  for this process; in fact, by substituting Eqs. (\ref{20})-(\ref{23})
into Eq. (\ref{13}) we find:
\be  \label{24}
\tau \; = \; \frac{1}{\omega} \, \frac{2 k_x^2}{\chi k} ~~~.
\ee
Furthermore, using the obvious relations:
\bea
k_x &=& k_1 \, \sin \theta \; = \; \frac{\omega}{c} \, n_1  \, \sin \theta
\label{25} \\
k &=& k_1 \, \cos \theta \; = \; \frac{\omega}{c} \, n_1  \, \cos \theta
\label{26} \\
\chi &=& \sqrt{k_1^2 \, \sin^2 \theta \, - \, k_2^2} \; = \;
\frac{\omega}{c} \, \sqrt{n_1^2 \, \sin^2 \theta \, - \, n_2^2}
\label{27} ~~~,
\eea
we finally get:
\be  \label{28}
\tau \; = \; \frac{1}{\nu} \, \frac{n_1 \, \sin^2 \theta}{\pi \, \cos
\theta \, \sqrt{n_1^2 \, \sin^2 \theta \, - \, n_2^2}} ~~~.
\ee
This formula can be directly checked with experiments. However, we firstly
observe the interesting feature of this expression which does satisfy the
property pointed out by Haibel and Nimtz \cite{haibel}. In fact, the time
$\tau$ in Eq. (\ref{28}) is just given, apart from a numerical factor
depending on the geometry and construction of the considered experiment, by
the reciprocal of the frequency of the radiation used. In a certain sense,
the numerical factor can be regarded as a ``correction'' factor to the
``universality'' property of Haibel and Nimtz. \\ Several experiments
measuring the tunnelling time in the considered process have been performed
\cite{ftir}. \\ In the experiment carried out by Balcou and Dutriaux
\cite{ftir}, two fused silica prisms with $n_1 = 1.403$ and an air gap
($n_2 = 1$) are used. They employed a gaussian laser beam of wave-length
$3.39 \, \mu m$ with an incident angle $\theta = 45.5^o$. Using these
values into Eq. (\ref{28}) we predict a tunnelling time of $36.8 \, fs$, to
be compared with the experimental result of about $40 \, fs$. As we can
see, the agreement is good and the ``correction'' factor in (\ref{28}) is
quite important for this to occur (compare with the Haibel and Nimtz
prediction of $11.3 \, fs$). \\ In the measurements by Mugnai, Ranfagni and
Ronchi \cite{ftir}, the microwave region is explored, with a signal whose
frequency is in the range $9 \div 10 \, GHz$. They used two paraffin prisms
($n_1 = 1.49$) with an air gap ($n_2 = 1$), while the incidence angle is
about $60^o$. For this experiment we predict a tunnelling time of $87.2 \,
ps$, while the experimental result is $87 {\pm} 7 \, ps$ \footnote{Note that
the value of $134 \, ps$ used by Haibel and Nimtz refers to the gap filled
with paraffin. In this case no tunnelling effect is present. We observe
that also for this experiment the ``correction'' factor in (\ref{28}) plays
a crucial role for the {\it tunnelling} times}. \\ Finally, we consider the
recent experiment performed by Haibel and Nimtz \cite{haibel} with a
microwave radiation at $\nu = 8.45 \, GHz$ and two perspex prisms ($n_1 =
1.605$) separated by an air gap ($n_2 = 1$). For an incident angle of
$45^o$, from (\ref{28}) we predict $\tau = 80.8 \, ps$. The observed
experimental result is, instead, $117 {\pm} 10 \, ps$. In this case, the
agreement is not very good (while, dropping the ``correction'' factor,
Haibel and Nimtz find a better agreement); probably this is due to the fact
that the condition of opaque barrier is not completely fulfilled.
\begin{figure}[t]
\epsfysize=1.5cm
\epsfxsize=7cm
\centerline{\epsffile{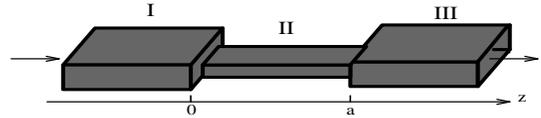}}
\caption{A waveguide with an undersized region.}
\label{und}
\end{figure}

\section{Undersized waveguide}

\noindent
Let us now consider propagation through undersized rectangular waveguides
as observed in \cite{under}. Also in this case, evanescent propagation
proceeds along one direction (say $z$) and the results obtained in Sect.
II may apply. With reference to Figure \ref{und}, a signal propagating inside
a ``large'' waveguide at a certain point undergoes through a ``smaller''
waveguide for a given distance $a$. As well known \cite{jackson}, the
signal propagation inside a waveguide is allowed only for frequencies
higher than a typical value (cutoff frequency) depending on the geometry of
the waveguide. In the considered setup, the two differently sized
waveguides I (or III) and II have, then, different cutoff frequencies
(the first one, $\omega_1$, is smaller than the second one, $\omega_2$),
and we consider the propagation of a signal whose frequency (or range of
frequencies) is larger than $\omega_1$ but smaller than $\omega_2$:
$\omega_1 < \omega < \omega_2$. In such a case, in the region $0 < z < a$
only evanescent propagation is allowed and, thus, the undersized waveguide
acts as a barrier for the photonic signal. With the same notation of Sect.
II, the dispersion relations in the large and small waveguide are,
respectively:
\bea
c \, k &=& \sqrt{\omega^2 \, - \, \omega_1^2}  \label{29} \\
c \, \chi &=& \sqrt{\omega_2^2 \, - \, \omega^2}  \label{30} ~~~,
\eea
so that:
\be
\frac{k}{\chi} \; = \; \sqrt{\frac{\omega^2 \, - \,
\omega_1^2}{\omega_2^2 \, - \, \omega^2}}  ~~~,
\ee
By substituting this expression into Eq. (\ref{11}), we immediately find
the tunnelling time in the regime of opaque barrier ($\chi a \gg 1$):
\be   \label{31}
\tau \; = \; \frac{1}{\nu} \, {\cdot} \, \frac{1}{\pi} \,
\sqrt{\frac{\nu^4}{(\nu^2 - \nu_1^2)(\nu_2^2 - \nu^2)}} ~~~.
\ee
On the contrary to what happens for tunnelling in total internal reflection
setups, the coefficient of the term $1/\nu$ isn't constant but depends
itself on frequency. Thus, in the case of undersized waveguides, the
assumed ``universality'' property of Haibel and Nimtz cannot apply in
general; depending on the cutoff frequencies, it is only a partial
approximate property for frequencies far way from the cutoff values (i.e.
when the term in the square root does not strongly depend on $\nu$). \\ Let
us now compare the prediction (\ref{31}) with the experimental results
obtained in \cite{under}. In the performed experiment we have microwave
radiation along waveguides whose cutoff frequencies are $\nu_1 = 6.56 \,
GHz$ and $\nu_2 = 9.49 \, GHz$, respectively. The radiation frequencies are
around $\nu = 8.7 \, GHz$, so that tunnelling phenomena occur in the
undersized waveguide. By substituting these values into Eq. (\ref{31}), we
predict a tunnelling time of $128 \, ps$, confronting the observed time of
about $130 \, ps$. \\ As it is evident, also for an undersized waveguide
setup the theory matches quite well with experiments. Note that, despite of
the rich frequency dependence in Eq. (\ref{31}), the Haibel and Nimtz
property also works quite well (although some correction needs),  since the
central frequency value of the radiation used in the experiment is far
enough from the cutoff values.
\begin{figure}[t]
\epsfysize=3.5cm
\epsfxsize=6cm
\centerline{\epsffile{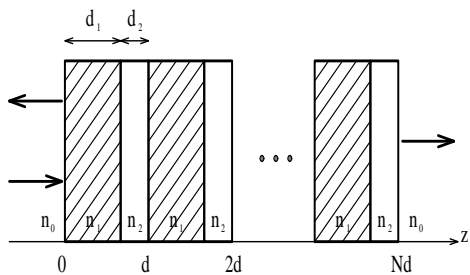}}
\caption{An ideal photonic bandgap device.}
\label{bandgap}
\end{figure}

\section{Photonic bandgap}

\noindent
The last phenomenon we consider is that of light propagation through
photonic bandgap materials. The ideal setup is depicted in Figure
\ref{bandgap}. Light impinges on a succession of thin plane-parallel films
composed of $N$ two-layer unit cells of thicknesses $d_1,d_2$ and constant,
real refractive indices $n_1,n_2$, embedded into a medium of index $n_0$.
It is known \cite{pbgthgen} that such a multilayer dielectric mirror
possesses a (one-dimensional) ``photonic bandgap'', that is a range of
frequencies corresponding to pure imaginary values of the wave-vector. In
practice, it is the optical analog of crystalline solids possessing
bandgaps. Increasing the number of periods will result in an exponential
increase of the reflectivity, and thus the opaque barrier condition can be
fulfilled. In general, the study of electromagnetic properties of such
materials is very complicated, and the dispersion relation we need to
evaluate the phase time in the proposed formalism is quite involved for
physical situations. This study was performed analytically in
\cite{scalora} where the dispersion relation (and other useful quantities)
was derived starting from the complex transmission coefficient of the
considered barrier. It is, then, quite a meaningless issue to get the
tunnelling time from the dispersion relation obtained from the transmission
coefficient, while it is easier to have directly the phase time $\tau$ from
Eq. (\ref{1}), where $\phi$ is the phase of the complex transmission
coefficient.

\subsection{Quarter-wave stack}

We first consider the relevant case in which each layer is designed so that
the optical path is exactly $1/4$ of some reference wave-length
$\lambda_0$: $n_1 d_1 = n_2 d_2 = \lambda_0/4$. In such a case, $\lambda_0$
corresponds to the midgap frequency $\omega_0$ ($\lambda_0=2
\pi c/\omega_0$). This condition is fulfilled in the considered experiments
\cite{pbg}. Finally, we further assume normal incidence of the light on the
photonic bandgap material.
\\
From \cite{scalora} we then obtain the following expression for the
transmission coefficient:
\be \label{tcoeff}
T \; = \; \left[ \left(A \, C \, -\, B \right) \, + \, i \, A \, D
\right]^{-1} ~~~,
\ee
where $A,B,C,D$ are real quantities given by:
\bea
A &=& \frac{\sin N \beta}{\sin \beta} \label{aa}
\\ B &=& \frac{\sin (N-1) \beta}{\sin \beta} \label{bb}
\\ C &=& a \, \cos \frac{\pi \omega}{\omega_0} \, + \, b
\\ D &=& c \, \sin \frac{\pi \omega}{\omega_0}
\eea
\bea
a &=& \frac{1 - r_{02}^2}{t_{02} t_{21} t_{12}} \\
b &=& \frac{r_{12}^2 (r_{02}^2 - 1)}{t_{02} t_{21} t_{12}} \\
c &=& \frac{2 r_{02} r_{12} - r_{02}^2 - 1}{t_{02} t_{21} t_{12}}
\eea
\bea
r_{ij} &=& \frac{n_i - n_j}{n_i + n_j} \\
t_{ij} &=& \frac{2 n_j}{n_i + n_j}
\eea
\be
\sin \beta \; = \; \frac{1}{t_{12} t_{21}} \, \sqrt{2 r_{12}^2 \, \left(
\cos \frac{\pi \omega}{\omega_0} \, - \, 1 \right) \, + \, \sin^2 \frac{\pi
\omega}{\omega_0}}
\ee
($i,j=1,2$).The phase $\phi$ of the transmission coefficient thus
satisfies:
\be \label{phi}
\tan \phi \; = \; \frac{A \, D}{B \, - \, A \, C} ~~~.
\ee
By substituting into Eq. (\ref{1}), we finally get an analytic expression
for the tunnelling time of light with frequency $\nu$ close to the midgap
one $\nu_0$ for $N$ layers:
\be  \label{32}
\tau \; = \; \frac{1}{\nu_0} \, {\cdot} \, \frac{1}{2} \, \frac{c \, \sinh N
\theta}{\sinh (N-1) \theta \, + \, (b-a) \, \sinh N \theta} ~~~,
\ee
where $\theta$ is simply obtained from:
\be  \label{33}
\sinh \theta \; = \; \frac{1}{2} \, \left( \frac{n_2}{n_1} \, - \,
\frac{n_1}{n_2} \right) ~~~.
\ee
Note that, although the tunnelling behaviour is quite different if the
number of periods $N$ is an even or odd number (see, for example,
\cite{diffe}), the expression for the tunnelling time given in (\ref{32})
(and also in (\ref{34})) is the same in both cases.
\\ For future reference, we also report the appropriate
formula for $N=k+(1/2)$ (integer $k$)
multilayer dielectric mirrors. In practice, this models the case of a
stratified medium whose structure has the form $n_1 n_2 n_1 n_2 \dots n_1
n_2 n_1$ (note, however, this is an approximation since, in general, $d/2$
is not equal to $a$). In such a case, Eq. (\ref{32}) is just replaced by:
\be  \label{34}
\tau \; = \; \frac{1}{\nu_0} \, {\cdot} \, \frac{1}{2} \, \frac{c \, \cosh N
\theta}{\cosh (N-1) \theta \, + \, (b-a) \, \cosh N \theta} ~~~,
\ee
Let us observe that, similarly to total internal reflection, at midgap the
time $\tau$ in Eq. (\ref{32}) or (\ref{34}) is again given by the
reciprocal of the frequency times a ``correction'' constant factor.
\\
We now analyze experimental results \cite{pbg} in the light of our
theoretical speculations. \\ In the experiment performed by Steinberg,
Kwiat and Chiao, the authors used a quarter-wave multilayer dielectric
mirror with a $(HL)^5H$ structure with a total thickness of $d=1.1 \mu m$
attached on one side of a substrate and immersed in air. Here, $H$
represents a titanium oxide film with $n_1=2.22$, while $L$ is a fused
silica layer with $n_2=1.41$. Thus, we have approximately $N=5+(1/2)$. As
incident light, they employed a wave-packet centred at a wave-length
$\lambda_0=702 nm$, corresponding to the midgap frequency $\nu_0$ of about
$427 THz$. By substituting these numbers in our formula (\ref{34}) we
predict a tunnelling time $\tau = 2.66 \, fs$, corresponding to a delay
time $\Delta t$, with respect to non tunnelling photons propagating at the
speed of light the distance $d$, of $-1.01 \, fs$. This has to be compared
with the experimental result of $\Delta t = -(1.47 {\pm} 0.2) \, fs$. However,
we point out that our analytical prediction is affected by two major
approximations. The first one is, as already remarked, that the
experimental sample is not really a $5+(1/2)$ periodic structure. A better
approximation is achieved by using Eq. (\ref{32}) with $N=6$ and
subtracting the time required for travelling at the speed of light the
quarter-wave thickness $d_2 = \lambda_0 /4 n_2$. In this case we have $\tau
= 2.02 \, fs$ or a delay time $\Delta t = -1.65 \, fs$, which is in better
agreement with the experimental result. Furthermore, in our analysis
(leading to Eq, (\ref{32}) or (\ref{34})) there is no room for considering
an asymmetric structure (like the substrate-air one) in which the photonic
bandgap material is embedded. This cannot be taken into account in an
analytic framework, but has to be studied using numerical matrix transfer
method which would give quite a good agreement with observations
\cite{scalora}.
\\
Finally, we consider the experiment carried out by Spielmann et al.
\cite{pbg} on alternated quarter-wave layers of fused silica $L$ and
titanium dioxide $H$ having the structure of $(substrate)(HL)^n(air)$ with
$N=3,5,7,9,11$. They used optical pulses of frequency $375$$THz$
corresponding to the midgap frequency of their photonic bandgap material.
Obviously, increasing $N$ we have a better realization of opaque barrier
condition. From Eq. (\ref{32}) with $N=11$ (note, however, that for $N\geq
5$ the factor $\sinh (N-1) \theta / \sinh N \theta$ is almost constant) we
have a tunnelling time of $2.98 \, fs$ to be compared with the observed
value of about $2.71 \, fs$. We address the fact that, apart the presence
of the asymmetric substrate-air structure which introduces some
approximation as discussed above, in the considered experiment the
incidence of the light on the sample is not normal, being $\approx 20^o$
the angle between the axis of the sample and the beam propagation
direction. In this case, the described computations are only approximated
ones and, again, the exact result can be obtained only through numerical
implementation. Nevertheless, also within the limits of our calculations,
the agreement between theory and experiment is quite good.
\\
A final comment regards the predictions of the ``universality'' property
proposed by Haibel and Nimtz. Neglecting the ``correction'' factor in Eq.
(\ref{32}) would yield the values of $\Delta t = -1.33 \, fs$ and $\tau =
2.67 \, fs$ for the delay time in the Steinberg, Kwiat and Chiao experiment
and the transit time for the Spielmann et al. experiment, respectively. In
both cases, the agreement with the observed values seems better than our
approximated predictions, showing that the presence of an asymmetric
substrate-air structure (and the non normal incidence in the second
experiment) pushes up the ``correction'' factor in Eq. (\ref{32}).

\subsection{Distributed Bragg Reflector}

We now relax the assumption of a quarter-wave stack $n_1 d_1 = n_2 d_2 =
\lambda_0/4$ but, for simplicity, we consider only the case in which
the photonic bandgap structure is embedded into a material whose refractive
index $n_0$ is equal to that of one of the two layers of the periodic
structure, that is $n_0 = n_2$. We again assume normal incidence of the
light on the photonic bandgap material. In this case the transmission
coefficient $T$ and its phase $\phi$ have the expressions as in
(\ref{tcoeff}) and (\ref{phi}), where $A$,$B$ are given by (\ref{aa}),
(\ref{bb}) and \cite{scalora}:
\bea
C &=& a \, \cos \pi \, \Omega_+ \, \omega \; - \; b \, \cos \pi \, \Omega_-
\, \omega
\\ D &=& - \, a \, \sin \pi \, \Omega_+ \, \omega \; + \; b \, \sin \pi \, \Omega_-
\, \omega
\eea
\be
\Omega_{\pm} \; = \; \frac{n_1 d_1 \, {\pm} \, n_2 d_2}{c}
\ee
\be
\sin \beta \; = \; \frac{1}{t_{12} t_{21}} \, \sqrt{P \, + \, Q \, + \, R}
\ee
\bea
   P &=& r_{12}^4 \, \sin^2 \pi \, \Omega_- \, \omega
\\ Q &=& 2 r_{12}^2 \, \left( \cos \pi \, \Omega_+ \, \omega \, \cos \pi \,
\Omega_- \, \omega \, - \, 1 \right)
\\ R &=& \sin^2 \pi \, \Omega_+ \, \omega
\eea
By substituting into Eq. (\ref{1}) we obtain the tunnelling time relative
to an $N$-layer structure:
\be
\tau \; = \; \frac{1}{\nu} \, \frac{X - Y}{Z}
\label{taubra}
\ee
\bea
   X &=& F \sin^2 \beta \cos N \beta \sin N \beta
\\ Y &=& G \left( \cos \beta \cos N \beta \sin N \beta - N \sin \beta \right)
\\ Z &=& 2 \, \sin \beta \left( D^2 \sin^2 N \beta + \sin^2 \beta \cos^2 N \beta \right)
\\ F &=& a \, \Omega_+ \, \omega \, \cos \pi \Omega_+ \omega \, - \, b \,
\Omega_- \, \omega \, \cos \pi \Omega_- \omega
\\ G &=& a^2 \, \Omega_+ \, \omega \, \sin^2 \pi \Omega_+ \omega \, + \, b^2 \,
\Omega_- \, \omega \, \sin^2 \pi \Omega_- \omega \, + \nonumber  \\
&~& - \, 2 a b \, (\Omega_+ + \Omega_-) \, \omega \, \sin \pi \Omega_+
\omega \, \sin^2 \pi \Omega_- \omega
\eea
Note that, again, the formula above for $\tau$ holds both for even $N$ and
for odd $N$.
\\ The obtained expression for the tunnelling time can be directly tested by analyzing
the experiment carried out by Mojahedi, Schamiloglu, Hegeler and Malloy
\cite{bragg2}. In this experiment the authors used a (1D) photonic crystal
composed of 5 polycarbonate sheets with refractive index $n_1=1.66$ and
thickness $d_1=1.27 \, cm$ separated by regions of air $n_2=1$ with
thickness $d_2=4.1 \, cm$. The bandgap was tuned to the main frequency
component ($\nu = 9.68 \, GHz$) of the incident microwave pulse. By
measuring both the signal travelling through the photonic bandgap structure
and the one propagating in free space, the authors found that the pulse
undergoing tunnelling has a delay time $\Delta t = - (440 {\pm} 20) \, ps$ with
respect to the other signal. By using Eq. (\ref{taubra}) with the above
numbers we predict a tunnelling time of $320 \, ps$ \footnote{Such a result
was also obtained in \cite{bragg2} using a formalism described in
\cite{bragg1} which is different from the one proposed here.},
corresponding to a delay time of $\Delta t = - 438 \, ps$, which is in
excellent agreement with the reported experimental result.
\\ We point out that, in this case, the simple $1/\nu$ law proposed by Haibel and
Nimtz does not work, since it would predict a tunnelling time $\tau = 103
\, ps$ or $\Delta t = -655 \, ps$. This can be easily explained by looking at
Eq. (\ref{taubra}). In fact, we immediately recognize that the
``correction'' factor in this equation is strongly frequency-dependent and,
for the frequency of the light used in the considered experiment, it is
sensibly bigger than one.

\section{Conclusions}

\noindent
In this paper we have scrutinized the recently proposed \cite{haibel}
``universality'' property of the photonic tunnelling time, according to
which the barrier traversal time for photons propagating through an
evanescent region is approximately given by the reciprocal of the photonic
frequency, irrespective of the particular setup employed. To this end, the
transit time in the relevant region, defined here as in Eq. (\ref{1}),
needs to be computed for the different explored phenomena, and in Sect. II
we have given general expressions for this time in the opaque barrier
limit. The peculiarities of a given photonic setup enter in these
expression only through the dispersion relation relating the wave-vector
and the frequency. More in detail, we have shown how the knowledge of the
ratio between the wave-vectors in the barrier region and outside it, as a
function of the photon frequency, is sufficient to evaluate the transit
time $\tau$ in Eq. (\ref{11}).
\\
Several specific cases, corresponding to the different classes of
experimentally investigated phenomena, have then been considered. In
particular, in Sect. III we have studied light propagation in a setup in
which the evanescent region is provided by total internal reflection, while
in Sect. IV the propagation through undersized waveguides has been
considered and, finally, in Sect. V the case of a photonic bandgap has been
analyzed. The relevant results for the three mentioned phenomena are given
in Eqs. (\ref{28}), (\ref{31}) and (\ref{32}) or (\ref{taubra}),
respectively. As can be easily seen from these expressions, the frequency
dependence of the tunnelling time for the cases of total internal
reflection and quarter-wave photonic bandgap is just as predicted by the
property outlined by Haibel and Nimtz \cite{haibel}, although we have
derived a ``correction'' factor depending on the geometry and on the
intrinsic properties of the sample (this factor is not far from unity). On
the contrary, such a factor is frequency dependent for undersized
waveguides and distributed Bragg reflectors, revealing a more rich
dependence of $\tau$ on $\nu$ than the simple $1/\nu$ one (see Eq.
(\ref{31})). We can then conclude that the ``universality'' property of
Haibel and Nimtz is only an approximation, but it gives the right order of
magnitude for the tunnelling time. This conclusion holds also for
undersized waveguide propagation, provided that the photon frequency is far
enough from the cutoff frequencies. \\
\begin{table}
\label{t1}
\caption{Comparison between predicted and observed tunnelling times for
several experiment (FTIR, UWG and PBG stands for frustrated total internal
reflection, undersized waveguide and photonic bandgap, respectively).
$\tau_{exp}$ is the experimental result while $\tau_{th}$ is our prediction
as from Eqs. (\ref{28}), (\ref{31}) and (\ref{32}) or (\ref{taubra}). For
reference to the Haibel and Nimtz property, we also report the value
$1/\nu$.}
\centering
\begin{tabular}{llccc}
\hline
\it Phenomenon & \it Experiment & $1/\nu$ & $\tau_{th}$ & $\tau_{exp}$ \\
\hline
   FTIR  & Balcou et al. \cite{ftir}   & $11.3 \, fs$ & $36.8 \, fs$ & $\sim 40 \, fs$
\\ FTIR  & Mugnai et al  \cite{ftir}   & $100 \, ps$  & $87.2 \, ps$ & $87 {\pm} 7 \, ps$
\\ FTIR  & Haibel et al. \cite{haibel} & $120 \, ps$ & $81 \, ps$   & $117 {\pm} 10 \, ps$
\\ UWG   & Enders et al. \cite{under}  & $115 \, ps$  & $128 \, ps$ & $\sim 130 \, ps$
\\ PBG($\lambda_0/4$) & Steinberg et al. \cite{pbg} & $2.34 \, fs$ & $2.02 \, fs$ &
$2.20 {\pm} 0.2 \, fs$
\\ PBG($\lambda_0/4$) & Spielmann et al. \cite{pbg}  & $2.67 \, fs$ & $2.98 \, fs$ &
$\sim 2.71 \, fs$
\\ PBG   & Mojahedi et al. \cite{bragg2} & $103 \, ps$  & $320 \, ps$ & $318 {\pm} 20 \, ps$
\\
\hline
\end{tabular}
\end{table}
We have then calculated the tunnelling times for the different existing
experiments and compared the theoretical values with the observed ones.
Results are summarized in Table 1, where we also report the Haibel and
Nimtz prediction $1/\nu$. From these we can see that, in general, the
agreement of our prediction with the experimental values is satisfactory.
As pointed out in the previous section, the calculations performed here for
photonic bandgap materials assume some approximations in treating the
complex sample, which are nevertheless required to obtain analytical
expressions. Our prediction then suffer of this and, in the case in which
the setup is designed to verify the quarter-wave condition $n_1 d_1 = n_2
d_2 = \lambda_0/4$, the simple $1/\nu$ rule fits better with experiments
while, for general photonic bandgap structures, the tunnelling time
displays a very complicated dependence on frequency. In this last case, as
well as in all other non photonic bandgap experiments, the ``correction''
factor introduced in this paper is quite relevant for the agreement with
observations to be good.


\acknowledgements

The author is indebted with Prof. E. Recami for many fruitful discussions
and useful informations about the subject of this paper and with an
anonymous referee for having drawn his attention on the experiments in
\cite{bragg1}, \cite{bragg2}. He also gratefully acknowledges Prof. A.
Della Selva for discussions.

\end{document}